\documentclass{svproc}
\usepackage{graphicx}
\usepackage{bm}
\usepackage{color}

\usepackage{url}

\begin{document}
\mainmatter              
\title{Visualization of Entanglement Geometry by Structural Optimization of Tree Tensor Network}
\titlerunning{Visualization of Entanglement Geometry by Structural Optimization of TTN}  
%
\author{Toshiya Hikihara\inst{1} \and Hiroshi Ueda\inst{2,3} \and Kouichi Okunishi\inst{4} \and Kenji Harada\inst{5} \and Tomotoshi Nishino\inst{6}}
\authorrunning{Toshiya Hikihara et al.} 
%
%
\institute{
Graduate School of Science and Technology, Gunma University, Kiryu, Gunma, Japan,\\
\and
Center for Quantum Information and Quantum Biology, Osaka University, Toyonaka, Japan, \\
\and
Computational Materials Science Research Team, RIKEN Center for Computational Science (R-CCS), Kobe, Japan, \\
\and
Department of Physics, Niigata University, Niigata, Japan, \\
\and
Graduate School of Informatics, Kyoto University, Kyoto, Japan, \\
\and
Department of Physics, Graduate School of Science, Kobe University, Kobe, Japan,}

\maketitle              

\begin{abstract}
In tensor-network analysis of quantum many-body systems, it is of crucial importance to employ a tensor network with a spatial structure suitable for representing the state of interest.
In the previous work [Hikihara et al., Phys. Rev. Research 5, 013031 (2023)], we proposed a structural optimization algorithm for tree-tensor networks.
In this paper, we apply the algorithm to the Rainbow-chain model, which has a product state of singlet pairs between spins separated by various distances as an approximate ground state.
We then demonstrate that the algorithm can successfully visualize the spatial pattern of spin-singlet pairs in the ground state.

\end{abstract}
\section{Introduction}

Tensor networks have proven to be an efficient format to represent low-energy states of quantum many-body systems and have been investigated intensively in various fields such as condensed-matter physics, quantum information, and quantum chemistry.\cite{Orus2019,OkunishiNU2022,Larsson2023}
In order to improve the accuracy of tensor networks in describing quantum states, it is known to be essential to employ the network with a spatial structure suitable for representing the target states efficiently.\cite{NishinoHOMAG2001,VerstraeteC2004,Vidal2007,MurgVLN2010,MurgVSNL2015,Larsson2019}

In the previous work\cite{HikiharaUOHN2023}, we focused on tree-tensor networks (TTNs), which are tensor networks without loops, and proposed a numerical algorithm to search for the optimal TTN structure for the precise representation of a target quantum state.
(We note that in the previous and present studies, we have considered the binary TTN.)
The algorithm is based on the guiding principle that the entanglement entropy (EE) at each auxiliary bond in the TTN should be as small as possible in order to minimize the truncation error.
We showed that the algorithm could successfully search out the optimal TTN structure for a quantum spin system whose interactions have a hierarchical spatial structure.
We have also found that the algorithm works effectively in improving the numerical accuracy of the TTN calculation.\cite{HikiharaUOHNunpub}

In this study, we apply the algorithm to a quantum spin system called the Rainbow chain, whose ground state is represented approximately by a product state of singlet pairs between spins separated by various distances.
We then show that from the optimal TTN structure and bond EEs obtained by the calculation, we can reproduce the pattern of the singlet-pair distribution of the ground state.
The result demonstrates that, while the TTN structural optimization algorithm is designed to minimize the truncation error, the algorithm can also visualize the entanglement geometry embedded in quantum many-body states.

The rest of this paper is organized as follows.
We discuss the guiding principle to define the optimal TTN structure in Sec.\ \ref{subsec:principle} and explain the core procedure of the proposed algorithm in Sec.\ \ref{subsec:algorithm}.
The Rainbow-chain model is introduced in Sec. \ref{subsec:model}, and our numerical results are presented in Sec. \ref{subsec:results}.
Section \ref{sec:summary} is devoted to a summary and concluding remarks.

\section{Guiding principle and algorithm}

\subsection{Least-entanglement-entropy principle}\label{subsec:principle}

Here, we discuss the guiding principle for determining the optimal structure of TTN.
A TTN represents a high-rank tensor such as a wave function of quantum many-body state as a product of local low-rank tensors.
If there is no restriction on the bond dimension, the dimension of legs of each tensor, the TTN can reproduce the wave function exactly.
In practice, however, we must set an upper bound on the bond dimension and truncate the Hilbert space to avoid the exponential growth of the computational cost on the system size.
This truncation causes an error in the representation of the target wave function.
The goal of the principle is to minimize this truncation error.

In a TTN, each bond has the corresponding bipartition of the system because of the property of TTN that it has no loop; see Fig.\ 1.
If the TTN can represent the entanglement of the bipartitions for all the bonds accurately enough, the TTN can precisely describe the target quantum state.
However, the maximum EE that a $\chi$-dimensional bond can carry is $\ln\chi$.
If the EE of a bipartition corresponding to a bond in a TTN exceeds this upper limit, the unrepresentable entanglement causes a loss of accuracy.
Conversely, if the EE at each bond in a TTN is sufficiently smaller than $\ln\chi$, the truncation error can be negligibly small.

\begin{figure}
\includegraphics[width=60mm]{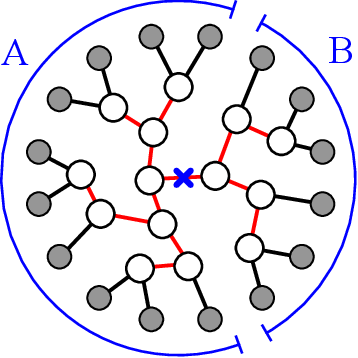}
\label{fig:TTN}
\caption{
Example of tree-tensor network.
The open circles denote tensors and the grey circles denote bare spins.
The red lines represent auxiliary bonds connecting tensors, while the block lines represent physical bonds connecting a tensor and a bare spin.
If one cuts the auxiliary bond indicated by a blue cross, the system is separated into two subsystems, A and B. 
}
\end{figure}

Here, it is worth noticing that, for a system with $N$ spins, the number of bonds in a TTN is $2N-3$, and determining the structure of TTN corresponds to selecting $(2N-3)$ bipartitions of the system from possible bipartitions of $2^{N-1}-1$.
From the arguments above, one can conclude that the structure of TTN which contains bonds with the smallest EEs is the ``optimal" one to represent the target quantum state.
The goal of the algorithm we proposed is to find the TTN structure that satisfies this least-EE principle.

Two comments are in order.
First, among the $(2N-3)$ bipartitions to be selected, the $N$ bipartitions corresponding to the physical bonds are fixed. We have a choice in selecting the remaining $(N-3)$ bipartitions corresponding to the auxiliary bonds. However, they can not selected independently due to the condition that the network indeed forms a TTN.
Second, for a practical realization of the least-EE principle, we have two options, i.e., minimizing the sum of EEs of each auxiliary bond in TTN or minimizing the maximum of the EEs.
In the proposed algorithm, we repeat a local reconnection of TTN to minimize the EE on a certain auxiliary bond, leaving EEs on the other bonds unchanged.
Therefore, the procedure simultaneously satisfies both options locally, while it is unclear which option is satisfied globally in the optimal TTN structure obtained.

\subsection{structural optimization algorithm}\label{subsec:algorithm}

The TTN structural optimization algorithm we proposed searches for the TTN structure that satisfies the least-EE principle discussed in the previous subsection.
Here, we describe the core process of the algorithm, i.e., the local reconnection of the network.
We refer the readers to Ref.\ \cite{HikiharaUOHN2023} for the entire algorithm and its details.

Let us focus on a part of TTN consisting of two adjacent tensors; see Fig.\ 2(a).
We regard the part as a single four-leg tensor, construct the effective Hamiltonian in the Hilbert space expanded by those four legs, and diagonalize the Hamiltonian to obtain the effective ground-state vector [Fig.\ 2(b)].
In the usual TTN variational method, one performs the singular-value decomposition to express the ground-state vector by a product of two tensors with the same structure as the original network.
The two tensors are thereby updated.
However, in performing the singular-value decomposition, there are two other candidates of local structure for splitting the ground-state vector into two tensors in addition to that of the original network structure, as shown in Fig.\ 2(c).
Furthermore, in each of those three candidates, the EE on the central bond takes a different value.
Therefore, we can adopt the local structure with the smallest EE on the central bond among the three candidates.
By this procedure, we can update the local connection of the network and the tensors simultaneously.
This is the core process of the proposed algorithm.
By sweeping the entire TTN with repeating the process, we can optimize the structure and the tensors of TTN.

\begin{figure}
\includegraphics[width=100mm]{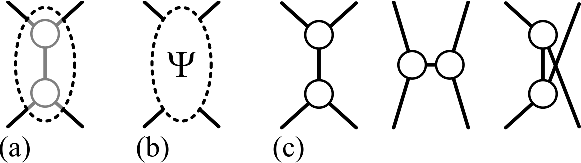}
\label{fig:reconnection}
\caption{
(a) The part to be updated in the TTN.
(b) The ground-state vector $\Psi$.
(c) Three candidates of local network structure to be adopted after the singular-value decomposition of $\Psi$.
}
\end{figure}

In TTN, any structure can be reached from an arbitrary initial TTN structure by repeating the local reconnection of the network.
Therefore, in principle, the proposed algorithm can search for the optimal structure from all possible structures of TTN.
However, in practice, it may occur that the optimization flow is trapped by a local minimum in the TTN-structural landscape and can not reach the global optimal structure.
A stochastic selection of the local connection may help overcome the difficulty.\cite{HikiharaUOHN2023}

\section{Results for Rainbow chain}

\subsection{Model}\label{subsec:model}

We apply the algorithm of TTN structural optimization to the one-dimensional Heisenberg model with position-dependent exchange constants which we call the Rainbow chain.
The Hamiltonian is given by
\begin{eqnarray}
\mathcal{H} = \sum_{j=1}^{N-1} J_j {\bm S}_j \cdot {\bm S}_{j+1},
\label{eq:Ham}
\end{eqnarray}
where ${\bm S}_j = \left\{ S^x_j, S^y_j, S^z_j \right\}$ is the spin-1/2 operator at $j$th site.
The exchange constants $J_j$ are set as
\begin{eqnarray}
J_j = \left\{
\begin{array}{ll}
1 & \left( j=\frac{N}{2} \right) \\
\alpha^{2n-1} & \left( j=\frac{N}{2} \pm n;~ n=1,...,\frac{N}{2}-1 \right).
\end{array} \right.
\label{eq:exchange_constant}
\end{eqnarray}
Namely, $J_{N/2}=1$ for the bond at the center of the chain, and $J_j$ decays as its position moves from the center to the edges of the chain [see Fig.\ 3(a)].
The parameter $\alpha$ ($0 < \alpha < 1$) controls the decay rate of the exchange constant.

\begin{figure}
\includegraphics[width=130mm]{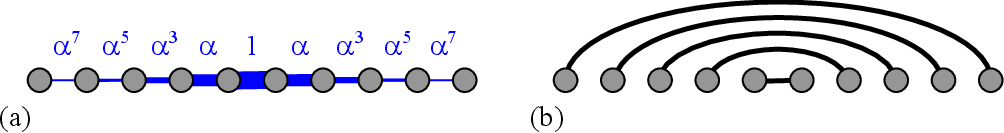}
\label{fig:Rainbow}
\caption{
(a) Exchange constants in the Rainbow chain with $N=10$.
(b) Singlet-pair distribution in the ground state of the Rainbow chain.
Solid lines represent the spin-singlet pairs.
}
\end{figure}

For a sufficiently small $\alpha$, the ground state of the Rainbow chain is obtained approximately by the perturbative real-space renormalization group approach\cite{MaDH1979,DasgupraM1980,XiL2008} as follows.
First, we focus on the center bond with the bond Hamiltonian $J_{N/2} {\bm S}_{N/2} \cdot {\bm S}_{N/2+1}$ with $J_{N/2}=1$, which is the strongest in the chain.
Since $J_{N/2}=1 \gg J_{N/2-1}=J_{N/2+1} = \alpha$, one can assume, to a good approximation, that the two spins ${\bm S}_{N/2}$ and ${\bm S}_{N/2+1}$ form a spin singlet in the ground state of the whole chain.
The two spins adjacent to the center bond, ${\bm S}_{N/2-1}$ and ${\bm S}_{N/2+2}$, are coupled via a second-order perturbative interaction $\tilde{J} {\bm S}_{N/2-1} \cdot {\bm S}_{N/2+2}$ with $\tilde{J}=(J_{N/2-1} J_{N/2+1})/(2 J_{N/2}) = \alpha^2/2$ and the Hamiltonian after the perturbation calculation recovers the same form as the original model of Eq.\ (\ref{eq:Ham}).
In the renormalized Hamiltonian, the strongest bond is the center one $\tilde{J} {\bm S}_{N/2-1} \cdot {\bm S}_{N/2+2}$ with $\tilde{J} = \alpha^2/2 \gg J_{N/2-2}=J_{N/2+2} = \alpha^3$ and one can repeat the same procedure above.
By iterating the renormalization procedure, one can eventually obtain the approximate ground state for the whole chain, in which two spins equidistant from the center of the chain form a spin-singlet pair, as depicted in Fig.\ 3(b).

\subsection{Numerical Results}\label{subsec:results}

We applied the structural optimization algorithm to the Rainbow chain [Eqs.\ (\ref{eq:Ham}) and (\ref{eq:exchange_constant})] with $\alpha=0.1$ and $N=8, 10$.
The calculation started from the matrix-product network shown in Fig.\ 4.
The truncation error of the calculation was negligibly small since, in the calculation, we kept all the states except the ones with an eigenvalue of the reduced density-matrix less than or equal to $1 \times 10^{-14}$; 
Indeed, the differences of the structural optimization calculation and the exact diagonalization in the ground-state energy (the correlation functions) were not more than $1 \times 10^{-14}$ ($1.3 \times 10^{-7}$).
In the following, we discuss the results for the chain with $N=10$, while the results for $N=8$ lead to essentially the same conclusion.

\begin{figure}
\includegraphics[width=70mm]{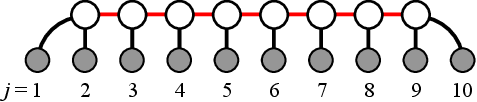}
\label{fig:initialMPN}
\caption{
Matrix-product network used as an initial network in the calculation.
The open circles denote the tensors and the grey circles denote the bare spins.
The black lines represent physical bonds directly connected to the bare spins.
The red lines represent auxiliary bonds.
}
\end{figure}

In order to confirm that the ground state obtained is indeed the Rainbow state depicted in Fig.\ 3(b), we have calculated the correlation functions in the ground state, 
\begin{eqnarray}
g(i,j) = \langle {\bm S}_i \cdot {\bm S}_j \rangle,
\label{eq:corr}
\end{eqnarray}
for all pairs $(i,j)$, where $\langle \cdots \rangle$ represent the expectation value in the ground state.
To analyze the quantum entanglement between the spins, we also calculate the concurrence, which is defined by\cite{HillW1997,Wootters1998,WangZ2002}
\begin{eqnarray}
\mathcal{C}(i,j) = \frac{2}{3} \max \left[
0, 2 |g(i,j)| - g(i,j) - \frac{3}{4}
\right],
\label{eq:concurrence}
\end{eqnarray}
where we have used the fact that the ground state is total spin singlet so that $\langle S^x_i S^x_j \rangle = \langle S^z_i S^z_j \rangle = g(i,j)/3$ and $\langle S^z_i \rangle = \langle S^z_j \rangle = 0$.
Note that the concurrence $\mathcal{C}(i,j) = 1$ if the spins ${\bm S}_i$ and ${\bm S}_j$ are maximally entangled in one of the Bell-pair states, while $\mathcal{C}(i,j) = 0$ if the two spins are in a separable state.
In Table \ref{tab:corr}, we present the results for five spin pairs having the largest concurrence.
For the other spin pairs $(i,j)$, the concurrence $\mathcal{C}(i,j)$ is zero, and the absolute value of the spin correlation, $|g(i,j)|$, is smaller than $0.186$.
These observations indicate that the ground state of the chain is approximately the Rainbow state, which consists of spin-singlet pairs of the pattern shown in Fig.\ 3(b) almost decoupled with each other.

\begin{table}
\caption{
Concurrence $\mathcal{C}(i,j)$ and spin correlation function $g(i,j)$ for the spin pairs $(i,j)$ that have the five largest $\mathcal{C}(i,j)$ among all pairs.
}
\label{tab:corr}
\begin{center}
\begin{tabular}{ccc}
\hline
$(i,j)$ & $\mathcal{C}(i,j)$ & $g(i,j)$ \\
\hline
 $(5, 6)$  & $0.982$  & $-0.741$  \\
 $(3, 8)$  & $0.926$  & $-0.713$  \\
 $(4, 7)$  & $0.923$  & $-0.712$  \\
 $(1,10)$  & $0.923$  & $-0.711$  \\
 $(2, 9)$  & $0.909$  & $-0.704$  \\
\hline
\end{tabular}
\end{center}
\end{table}

Figure\ 5 presents the optimal TTN and the bond EE $\mathcal{S}$ obtained by the TTN structural optimization.
The EEs on the physical bonds directly connected to the bare spins take the value $\mathcal{S} = \ln 2 = 0.693...$, reflecting the fact that the ground state is a total spin-singlet.
Two spins forming a spin-singlet pair in the Rainbow state are connected via only a single tensor in the optimal TTN.
Furthermore, those blocks of singlet pairs are connected by auxiliary bonds carrying a small EE, $\mathcal{S} \le 0.24$.
Here, we emphasize that the appearance of a bond with the small EE indicates that the subsystems connected by the bond are almost decoupled.
We can thereby conclude from the optimal TTN with the values of EE on each bond that the ground state is indeed the Rainbow state shown in Fig.\ 3(b).

\begin{figure}
\includegraphics[width=130mm]{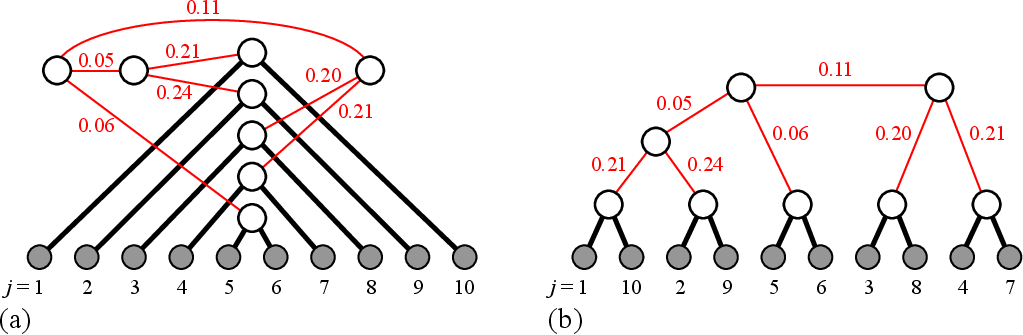}
\label{fig:optimalTTN}
\caption{
Optimal TTN obtained by the structural optimization algorithm for the Rainbow chain with $\alpha=0.1$ and $N=10$.
The open circles denote the tensors and the grey circles denote the bare spins.
In (a), the spins are arranged by the site number.
In (b), the spins are arranged so that two adjacent spins in the panel form the singlet pairs.
The numbers presented at the thin red lines represent the EE on the corresponding bond.
The thick black lines represent the physical bonds with EE of $\ln 2$.
}
\end{figure}

\section{Summary and Discussions}\label{sec:summary}

In summary, we have studied the ground state of the Rainbow-chain model [Eqs.\ (\ref{eq:Ham}) and (\ref{eq:exchange_constant})] using the TTN structural optimization algorithm proposed in Ref.\ \cite{HikiharaUOHN2023}.
It has been shown that from the optimal TTN structure and the results of EE on each bond obtained by the calculation, we can determine that the ground state consists of spin-singlet pairs almost decoupled from each other and distributed in the pattern depicted in Fig.\ 3(b).

The results of the present study demonstrate that the TTN structural optimization algorithm is able to visualize the singlet-pair distribution for a product state containing entangled spin pairs of not only neighboring spins but also distant spins.
This observation suggests that the algorithm may be useful for analyzing such a type of state, including the random-singlet state realized in the antiferromagnetic random spin chains.\cite{MaDH1979,DasgupraM1980,Fisher1994}
However, here, we must comment that in a state containing long-range singlet pairs such as the random-singlet state, the excitation energy to break the singlet pair usually decays exponentially with the distance between two spins and often becomes smaller than the numerical resolution of diagonalization calculation of the effective Hamiltonian to obtain the ground state.
If such a situation occurs, the TTN structural optimization is also expected to fail to visualize the singlet-pair distribution correctly since the diagonalization calculation can not reproduce the accurate ground state.
Indeed, for the Rainbow chain, we have found that for $\alpha=0.1$ and $N=12$, the spin correlation between the edge spins can not reproduced correctly even by the exact diagonaliation with double-precision numbers.
We emphasize that this is not a problem of TTN structural optimization but a problem of the method to obtain the ground state, and we must employ a method that can accurately compute the ground state with such a hierarchical energy structure.
While using quadruple-precision numbers could be an immediate measure, a more fundamental solution may be a renormalization-group approach such as the tensor-network strong-disorder renormalization-group method\cite{HikiharaFS1999,GoldsboroughR2014,SekiHO2020,SekiHO2021}.

While we have shown in the present study that the proposed algorithm can clarify the entanglement geometry of quantum states, we have also found that the algorithm can be practically effective in improving the accuracy of numerical calculations for various quantities, including the variational ground-state energy and correlation functions.
The results will be reported elsewhere.\cite{HikiharaUOHNunpub}

The present algorithm optimizes the TTN structure and the tensors simultaneously to obtain the ground-state wave function of a given model.
In the algorithm, the process of diagonalizing the effective Hamiltonian to compute the ground-state wave function can be replaced with a process of constructing the wave function from the previously obtained tensors.
Such a replacement leads to a revised algorithm that can generate an accurate representation of a given arbitrary high-rank tensor in the TTN format with the optimal network structure.\cite{WatanabeUH2024}
The revised algorithm is expected to be useful in a wide range of fields that utilize the tensor-network representation of a high-rank tensor.

\vspace{0.6cm}
{\it Acknowledgments.} This work is partially supported by KAKENHI Grant Numbers JP22H01171, JP21H04446, JP21K03403, JP20K03766, and Grants-in-Aid for Transformative Research Areas "The Natural Laws of Extreme Universe---A New Paradigm for Spacetime and Matter from Quantum Information" (KAKENHI Grant No. JP21H05182 and No. JP21H05191) from MEXT of Japan. H.U. acknowledges support from MEXT Q-LEAP Grant No. JPMXS0120319794, JST COI-NEXT No. JPMJPF2014. H.U. and T.N. are supported by the COE research grant in computational science from Hyogo Prefecture and Kobe City through Foundation for Computational Science.


%

\end{document}